\pdfoutput=1  

\documentclass[format=acmsmall, nonacm]{acmart}

\usepackage{mathtools}
\usepackage{cleveref}
\usepackage{xfrac}
\usepackage{adjustbox}
\usepackage{stmaryrd}    
\usepackage{hyperref}
\hypersetup{
  colorlinks = true,
  allcolors=.
}
\usepackage{hhline}

\usepackage{tikz}
\usetikzlibrary{cd}

\DeclareMathAlphabet{\mathpzc}{OT1}{pzc}{m}{it} 

\usepackage{enumitem} \setlist{nosep}

\definecolor{darkblue}{rgb}{0.05,0.25,0.65}
\definecolor{greenii}{RGB}{20,140,10}
\definecolor{darkgreen}{rgb}{0.00,0.85,0.1}
\definecolor{lightgray}{rgb}{0.9,0.9,0.9}
\definecolor{orangeii}{RGB}{200,100,5}
\definecolor{darkyellow}{rgb}{.91,.91,0}

\newcommand{\ComplexNumbers}{\mathbb{C}}

\newcommand{\HomotopyQuotient}[2]{
  #1 \!\sslash\! #2
}

\usepackage[safe]{tipa} 
\newcommand{\shape}{
  \raisebox{1pt}{\rm\normalfont\textesh}
}

\newcommand{\TopologicalSpace}{\mathrm{X}}

\newcommand{\FredholmOperators}{
  \mathrm{Fred}
}

\newcommand{\ShiftedLevel}{\kappa}
\newcommand{\weight}{\mathrm{w}}

\newcommand{\ConfigurationSpace}[1]{  \underset{
    \scalebox{.65}{$
      \{1,\cdots,#1\}
    $}
  }
  {\mathrm{Conf}}
}

\newcommand{\TED}{TED}
\newcommand{\TQC}{TQC}

\newcommand{\DualTorus}[1]{\widehat{\mathbb{T}}{}^{#1}}

\newcommand{\SpecialUnitaryLieAlgebra}[1]{\mathfrak{su}_{#1}}

\newcommand{\suAffine}[2]{\widehat{\SpecialUnitaryLieAlgebra{#1}}^{\raisebox{-2.5pt}{\scalebox{.73}{\hspace{-1pt}$#2$}}}}

\newcommand{\suTwoAffine}[1]{\suAffine{2}{#1}}

\begin{document}

\title{
  Topological Quantum Programming in {\tt TED-K}
}

\author{Hisham Sati}
\email{hsati@nyu.edu}
\orcid{0002-7998-5206}
\affiliation{
  \institution{New York University}
  \city{Abu Dhabi}
  \streetaddress{P.O. BOX 129188}
  \country{UAE}
}

\author{Urs Schreiber}
\email{us13@nyu.edu}
\orcid{0002-2876-8877}
\affiliation{
  \institution{New York University}
  \city{Abu Dhabi}
  \streetaddress{P.O. BOX 129188}
  \country{UAE}
}

\begin{abstract}
  While the realization of scalable quantum computation will arguably require topological stabilization and, with it, topological-hardware-aware quantum programming and topological-quantum circuit verification, the proper combination of these strategies into dedicated {\it topological quantum programming} languages has not yet received attention.

\smallskip
  Here we describe a fundamental and natural scheme that we are developing, for typed functional (hence verifiable) topological quantum programming which is {\it topological-hardware aware} -- in that it natively reflects the universal fine technical detail of topological q-bits, namely of symmetry-protected (or enhanced) topologically ordered Laughlin-type anyon ground states in topological phases of quantum materials.

  \smallskip

  What makes this work is:

\vspace{1mm}
 \begin{itemize}[leftmargin=.4cm]

\item[{1.}] our recent result \cite{SS22AnyonictopologicalOrder}\cite{SS22AnyonicDefectBranes} that wavefunctions of realistic and technologically viable anyon species -- namely of $\mathfrak{su}(2)$-anyons such as the popular Majorana/Ising anyons but also of computationally universal Fibonacci anyons -- are reflected in the twisted equivariant differential (TED) K-cohomology of
  configuration spaces of codimension=2 nodal defects in the host material's crystallographic orbifold;

\item[{2.}] combined with our earlier observation
\cite{SS21EPB}\cite{SS20OrbifoldCohomology}\cite{Schreiber14}
that such TED generalized cohomology theories on orbifolds interpret intuitionistically-dependent linear data types in cohesive homotopy type theory (HoTT), supporting a powerful modern form of modal quantum logic.
\end{itemize}

  \vspace{1mm}
  Not only should this emulation of anyonic topological hardware functionality via {\tt TED-K} implemented in cohesive HoTT make advanced formal software verification tools available for hardware-aware topological quantum programming, but the constructive nature of type-checking a {\tt TED-K} quantum program in cohesive HoTT on a classical computer using existing software (such as {\tt Agda-$\flat$}) has the potential to amount at once to classically simulating the intended quantum computation at the deep level of physical topological q-bits.

  \smallskip
  This would make {\tt TED-K} in cohesive HoTT
  an ideal software laboratory for topological quantum computation on technologically viable types of topological q-bits, complete with ready compilation to topological quantum circuits as soon as the hardware becomes available.

  \smallskip
  In this short note we give an exposition of the basic ideas, a quick review of the underlying results and  a brief indication of the basic language constructs for anyon braiding via {\tt TED-K} in cohesive HoTT. The language system is under development at the \href{https://ncatlab.org/nlab/show/Center+for+Quantum+and+Topological+Systems}{\it Center for Quantum and Topological Systems} at the Research Institute of NYU, Abu Dhabi. For supplementary material to this announcement see:
  \href{https://ncatlab.org/schreiber/show/Topological+Quantum+Computation+in+TED-K}{\tt ncatlab.org/schreiber/show/TQCinTEDK}.

\end{abstract}


\maketitle

\medskip

\newpage

\noindent
{\bf Need for topological quantum programming.}
The key \cite{Shor95}
to making the idea of quantum computation
(e.g. \cite{BenentiCasatiRoissini18}\cite{NielsenChuang00}) a viable practical reality remains (e.g. \cite{Monz22}) the stabilization of quantum circuits against noise and decoherence ({\it fault-tolerance}, e.g. \cite{Preskill97a}\cite{Preskill97b}\cite{Gottesman07}). This may conceivably be done after the fact, via {\it quantum error correction} (\cite{Shor95}, see \cite{Terhal15}\cite{BallCentellesHuber20}), but optimally such errors would be avoided in the first place: The grand promise of {\it topological quantum computation} ({\TQC}, \cite{Kitaev03}\cite{FKLW01}\cite{NSSFS08}, review in \cite{Wang10}) is to utilize {\it topological effects}
in the underlying quantum materials
(see \cite{Stanescu20}\cite{ZCZW19}\cite{MoessnerMoore21}\cite{SS22AnyonictopologicalOrder})
to constrain the pathways along which quantum coherence can decay at all. It may be argued \cite{DasSarma22}\footnote{
``{\it The q-bit systems we have today are a tremendous scientific achievement, but they take us no closer to having a quantum computer that can solve a problem that anybody cares about. [\ldots] What is missing is the breakthrough [\ldots] bypassing quantum error correction by using far-more-stable q-bits, in an approach called topological quantum computing.}'' \cite{DasSarma22}.} that topological protection is not an option but a necessity for realizing useful quantum computation that deserves the name.

\medskip
Since the principal hardware component of TQC -- namely {\it anyonic} topological order in topological phases of quantum materials (\cite{Kitaev06}\cite{SS22AnyonictopologicalOrder}) -- has recently been demonstrated in experiment (\cite{BartolomeiEtAl20}\cite{NLGM20}\cite{MintairovEtAl21}, notably in a promising novel reciprocal incarnation via band nodes in momentum space \cite{BzdusekEtAl20}\cite[Rem. 3.9]{SS22AnyonictopologicalOrder}), there seems to be no fundamental technical obstruction against the eventual construction of TQC machines, ambitious as it may still be.
Hence, while the engineers are occupied with the task of constructing topological quantum hardware, theorists must become serious about the upcoming practice of {\it topological quantum  programming}.

\medskip
\noindent
{\bf Nature of topological quantum programming.}
Efficiency demands that a programming language be {\it hardware-aware}, in that its design principles align with the functionality that the machine offers. This has become common-place for available and near-future toy quantum computers (e.g. \cite{ShiEtAl20}\cite{ZhuCross20}),
but in view of the required
topological quantum revolution it remains to ask:

\begin{center}
{\it How can a quantum programming language be aware of {\rm topological} quantum hardware?}
\end{center}

By this we mean that language structures reflect the established implementation paradigm for topological quantum computation (see \cite{SS22AnyonictopologicalOrder} for pointers), including:

\smallskip

\begin{itemize}[leftmargin=.8cm]
\item[(1.)] topological q-bits encoded in topologically ordered ground states depending on given positions of $\suTwoAffine{k}$-anyon defects at any admissible Chern-Simons level $k$,

\item[(2.)] quantum logic gates operating by adiabatic braiding of these positions.
\end{itemize}

\smallskip

\noindent
But this question of topological-hardware awareness has not received much attention yet, apart for the abstract question of compiling quantum programs from braid gate circuits (\cite{BonesteelEtAl05}\cite{Hormozi07}).

\smallskip

 We present a {\it principal answer} to this question by
going to the very bottom of the concepts of:

\vspace{.1cm}

\hspace{-.6cm}
$
\left.
\mbox{
\begin{tabular}{ll}
(a)
&
{\bf computation}
\\
(b)
&
{\bf algebraic topology}
\\
(c)
&
{\bf quantum physics}
\end{tabular}
}
\right\}
\mbox{
\hspace{-.1cm}
in their guise of
\hspace{-.05cm}
}
\left\{
\mbox{
\begin{tabular}{lll}
(a)
&
{\bf homotopy type theory}
&
\cite{UFP13}\footnotemark
\\
(b)
&
{\bf generalized cohomology}
&
\cite{TamakiKono06}
\\
(c)
&
{\bf charge quantization}
&
\cite{FSS20Character},
\end{tabular}
}
\right.
$
\footnotetext{
  For further introduction to homotopy type theory see also
  \cite{Escardo19}\cite{BBCDG21}.
  For its semantics in $\infty$-topoi
  as originally envisioned in \cite[\S 3]{Awodey12}
  -- such as those considered in the references \cite{SS20OrbifoldCohomology}\cite{SS21EPB}\cite{SS22AnyonictopologicalOrder} to which we refer here --
  see \cite{Shulman19}, reviewed in \cite{Riehl22}.
}
\vspace{.1cm}

\noindent
where we find a novel programming scheme
which natively connects:

\smallskip
\begin{itemize}
\item[(1.)]
the mesoscopic physical principles of
{\bf topologically ordered quantum states}

\cite{Kitaev06}\cite{NSSFS08}
\item[(2.)]
the high-level language of quantum logic in the form of {\bf dependent linear type theory}

\cite{Schreiber14}\cite{Vakar15}\cite{FKRS20}\cite{RileyFinsterLicata21}\cite{Riley22}.
\end{itemize}

\medskip
\noindent
{\bf TED K-theory for topological quantum programming.}
The connective tissue between these concepts is
(\cite{SS22AnyonictopologicalOrder}\cite{SS22AnyonicDefectBranes})
the cohesive generalized cohomology theory called (\cite{SS21EPB}\cite{SS22TED} following \cite{AtiyahSegal04}\cite{HopkinsSinger05}):

\vspace{-.2cm}
\begin{center}
{\it Twisted, equivariant, differential K-theory
  (henceforth}  \TED-K-theory{\it)}.
\end{center}

\noindent
Namely, TED K-theory naturally and accurately reflects (we indicate in a moment how this works):
\begin{itemize}[leftmargin=.4cm]
\item
the fundamental {\bf principles of topological quantum physics} (as first highlighted in \cite{FreedMoore12}, see \cite[\S 1]{SS22AnyonictopologicalOrder} for more), in fact of anyonic topologically ordered quantum ground states (\cite[\S 2]{SS22AnyonictopologicalOrder})
\end{itemize}
\noindent

\begin{itemize}[leftmargin=.4cm]
\item
in the mathematics of
{\bf geometric} \cite{Brown73}
{\bf equivariant} \cite{tomDieck79}
{\bf stable} \cite{Adams74}
{\bf homotopy theory} \cite{Strom11}.
\end{itemize}

\smallskip

\noindent
But there exists a programming language for synthetic constructions in this rich form of homotopy theory, namely
 {\bf cohesive homotopy type theory} \cite{Schreiber13}\cite{Schreiber14}\cite{Schreiber14Talk}\cite{SchreiberShulman14}\cite{Schreiber15}\cite{Shulman15}\cite{Wellen18a}\cite{Corfield20}\cite{SS20OrbifoldCohomology}\cite{Myers21}\cite{RileyFinsterLicata21}\cite{SS21EPB}\cite{Riley22};
 see
 \cite{Licata13}\cite{Wellen18b}\cite{Shulman21}\cite[p. 5-6]{SS20OrbifoldCohomology} for exposition and further pointers.

\smallskip

This way,  {\TED}-K-theory {\it is} a natural topological quantum programming scheme when handled appropriately: Its implementation in cohesive HoTT makes it a programming language construct, and its reflection of anyonic topological quantum order then makes it a topological hardware-aware quantum programming language
(we illustrate this in a moment):

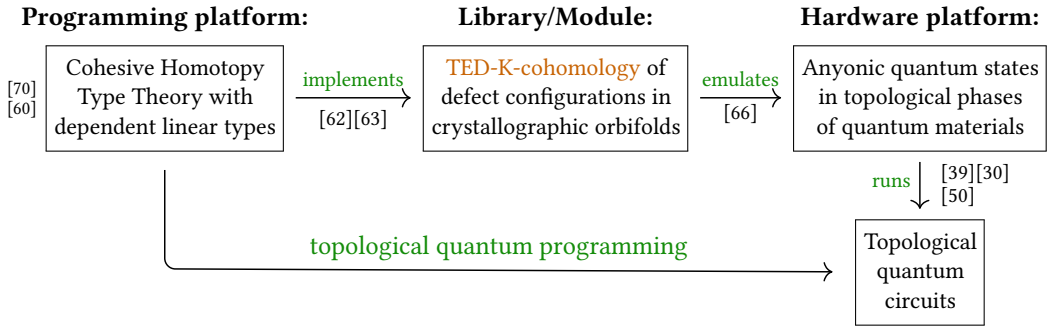
\vspace{-.5cm}

$$
  \begin{tikzcd}[
    column sep={between origins, 69pt}
    ]
    \mbox
    {
      \bf
      Programming platform:
    }
    &[+10pt]&
    \mbox
    {\bf Library/Module:}
    &&
    \mathclap{
    \mbox
    {\bf Hardware platform:}
    }
    \\[-20pt]
    \mathllap{
      \scalebox{.7}{
      \def\arraystretch{.9}
       \begin{tabular}{c}
         \cite{Schreiber14}
         \\
         \cite{RileyFinsterLicata21}
       \end{tabular}
      }
      \hspace{-2pt}
    }
    \fbox{
      \hspace{-10pt}
      \small
      \begin{tabular}{c}
        Cohesive Homotopy
        \\
        Type Theory with
        \\
        dependent linear types
      \end{tabular}
      \hspace{-10pt}
    }
    \ar[
      rr,
      "{
        \color{greenii} \footnotesize \bf
        \mbox{implements}
      }",
      "{
        \scalebox{.8}{
          \cite{SS20OrbifoldCohomology}\cite{SS21EPB}
        }
      }"{swap, yshift=-2pt}
    ]
    \ar[
      drrrr,
      rounded corners,
      to path={
           ([yshift=-3pt]\tikztostart.south)
       --  ([yshift=-40pt]\tikztostart.south)
       --
       node[above]
         {
           \scalebox{1}{
             \color{greenii}
             topological quantum programming
           }
         }
        ([xshift=-4pt]\tikztotarget.west)
      }
    ]
    &[+2pt]&
    \fbox{
      \hspace{-10pt}
      \small
      \begin{tabular}{c}
        {\color{orangeii}\TED-K-cohomology} of
        \\
        defect configurations in
        \\
        crystallographic orbifolds
      \end{tabular}
      \hspace{-10pt}
    }
    \ar[
      rr,
      "{
        \color{greenii} \footnotesize \bf
        \mbox{emulates}
      }"{yshift=2pt},
      "{
        \scalebox{.8}{
          \cite{SS22AnyonictopologicalOrder}
        }
      }"{swap}
    ]
    &[+2pt]&
    \fbox{
      \hspace{-10pt}
      \small
      \begin{tabular}{c}
        Anyonic quantum states
        \\
        in topological phases
        \\
        of quantum materials
      \end{tabular}
      \hspace{-10pt}
    }
    \ar[
      d,
      "{
        \color{greenii} \footnotesize \bf
        \mbox{runs}
      }"{xshift=-1pt, yshift=1pt, swap},
      "{
        \scalebox{.8}{
          \hspace{-.2cm}
          \def\arraystretch{.9}
          \begin{tabular}{l}
            \cite{Kitaev03}\cite{FKLW01}
            \\
            \cite{NSSFS08}
          \end{tabular}
        }
      }"
    ]
    \\[+0pt]
    &&
    &&
    \fbox{
      \hspace{-10pt}
      \small
      \begin{tabular}{c}
        Topological
        \\
        quantum
        \\
        circuits
      \end{tabular}
      \hspace{-10pt}
    }
  \end{tikzcd}
$$
\vspace{-.35cm}

To put this in perspective, notice that existing quantum programming languages
(surveyed in \cite{GGA21})
are, at their core, formal languages for (the category of) {\it linear algebra}
(as foreseen in \cite{Pratt92}\cite{AbramskyCoecke04}\cite{Selinger04}\cite{AbramskyDuncan06}\cite{Duncan06}), whose data types are {\it linear types} (such as Hilbert spaces) of quantum states and whose algorithms are linear maps (unitary operators) between these: quantum circuits \cite{DalLagoFaggian12}.

\smallskip

In topological refinement of this state of affairs, cohesive homotopy type theory is in particular
(\cite[\S 4.1]{Schreiber13}\cite{Schreiber14}\cite{Schreiber14Talk}\cite{RileyFinsterLicata21})
a language (specifically: a {\it functional language}, like {\tt QML} \cite{AltenkirchGrattage05} or {\tt Quipper} \cite{GLRSV13})
for {\it linear homotopy theory} traditionally known as {\it stable homotopy theory},
whose data types include ''linear homotopy types'' known as {\it spectra},
a prominent example of which is the spectrum $\mathrm{KU}$ representing topological K-theory (e.g. \cite[\S B.2]{AGP08}).

\vspace{-.2cm}
{\small
\begin{center}
\def\arraystretch{2}
\begin{tabular}{|c||c|c|}
\hline
&
\bf
Traditional quantum programming
&
\bf
{\tt TED-K} in cohesive HoTT
\\
\hline
\hline
\bf Type theory
&
\def\arraystretch{1}
\begin{tabular}{c}
  Linear
  type theory
  \\
  (linear algebra)
  \clap{$\phantom{\vert_{\vert_{\vert}}}$}
\end{tabular}
&
\def\arraystretch{1}
\begin{tabular}{c}
  Linear homotopy
  type theory
  \\
  (stable homotopy theory)
\end{tabular}
\\
\hline
\bf Data types
&
\def\arraystretch{1}
\begin{tabular}{c}
Linear types
\\
(vector spaces)
  \clap{$\phantom{\vert_{\vert_{\vert}}}$}
\end{tabular}
&
\def\arraystretch{1}
\begin{tabular}{c}
  Linear homotopy types
  \\
  (cohesive spectra)
\end{tabular}
\\
\hline
\def\arraystretch{.9}
\begin{tabular}{c}
  \bf
  Classical adiabatic
  \\
  \bf
   control
\end{tabular}
&
\def\arraystretch{1}
\begin{tabular}{c}
  Dependent linear types
  \\
  (vector bundles)
\end{tabular}
&
\def\arraystretch{1}
\begin{tabular}{c}
  Dependent linear homotopy types
  \\
  (cohesive parameterized spectra)
\end{tabular}
\\
\hline
\def\arraystretch{.9}
\begin{tabular}{c}
\bf   Unit of
  \\
\bf   information
\end{tabular}
&
\def\arraystretch{1}
\begin{tabular}{c}
  Q-bit in $\ComplexNumbers$
\end{tabular}
&
\def\arraystretch{1}
\begin{tabular}{c}
  Character in $\mathrm{KU}_G^\tau(-; \mathbb{C})$
\end{tabular}
\\
\hline
\end{tabular}
\end{center}
}
\vspace{+.05cm}

In fact, this subsumes {\it dependent} linear homotopy types \cite{Schreiber14}\cite{Vakar15}\cite{RileyFinsterLicata21}\cite{Riley22},
which encode {\it twisted} generalized cohomology theories (see e.g. \cite{GS-tAHSS}\cite[\S 2.2]{FSS20Character}), such as
\TED-K-theory. Computationally, the dependency and thus of linear data types on ordinary (i.e. ``intuitionistic'') data types reflects the
{\it controlling} of quantum computation by classical computers (in the spirit of \cite{PerdrixJorrand06}), specifically the {\it adiabatic quantum computation} (e.g. \cite{AlbashLidar16}) by ``slow'' variation of external parameters, of which anyon braiding is an example (e.g. \cite[pp. 6]{FKLW01}\cite[p. 6]{NSSFS08}\cite{CGDS11}\cite{CLBFN15}), to which we come back in a moment.

\medskip

\noindent
{\bf The topological quantum trilogy.}
To appreciate how natural this enhanced programming scheme actually is,
notice that we may understand the passage from the left to the right column in the above table as
{\it topologization} followed by  {\it quantization} of the classical
{\it computational trilogy}
({\bf I} below, due to \cite[\S 1]{Mellies06} following \cite{LambekScott86}, review in \cite[\S 3]{Eades12})
which puts into mutual relation the theories of \\
{\bf (i)} computation, {\bf (ii)} type theory {\bf (iii)} category theory:

\vspace{1mm}
\begin{itemize}[leftmargin=.6cm]
\item First (in {\bf II}) the {\it topological computational trilogy}
(see \cite{Shulman18} with \cite[pp. 5-6]{SS20OrbifoldCohomology})
identifies:
\\
{\bf (i)} dependent/contextual computation,
{\bf (ii)} homotopy type-theory, and
{\bf (iii)} homotopy theory;

\item
and then (in {\bf III})  the {\it quantum computational trilogy}
identifies
{\bf (i)} classically-controlled quantum computation,
{\bf (ii)} dependent linear homotopy type-theory, and
{\bf (iii)} algebraic topology.
\end{itemize}

\smallskip
\hspace{-.7cm}
\scalebox{.9}
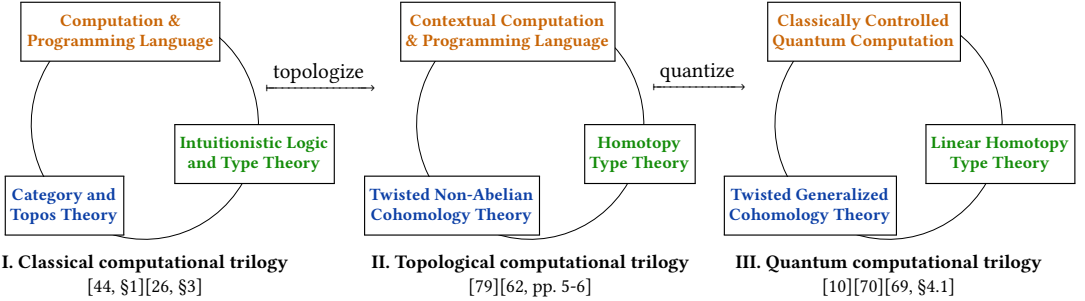
{
\begin{tikzpicture}
\hspace{-2mm}
\tiny

\begin{scope}[shift={(-5.7,0)}]
\begin{scope}[rotate=(60), scale=(.7)]

\draw (0,0) circle (2.4);

\draw
  (45:2)
    node
  {
  \scalebox{1.2}{
   {
    \fcolorbox{black}{white}{ \bf
    \hspace{-9pt}
    \color{orangeii}
    \begin{tabular}{c}
      Computation \&
      \\
      Programming Language
    \end{tabular}
    \hspace{-9pt}
    }
    }
  }};

\draw
  (45+120:2.4)
    node
  {
    \scalebox{1.2}{
    {
    \fcolorbox{black}{white}{\bf
    \hspace{-8pt}
    \color{darkblue}
    \begin{tabular}{c}
      Category and
      \\
      Topos Theory
    \end{tabular}
    \hspace{-9pt}
    }
    }
  }};

\draw
  (45+240:2.4)
    node
  {
   \scalebox{1.2}{
   {
    \fcolorbox{black}{white}{\bf
    \hspace{-8pt}
    \color{greenii}
    \begin{tabular}{c}
      Intuitionistic Logic
      \\
      and Type Theory
    \end{tabular}
    \hspace{-9pt}
    }
    }
  }};

\end{scope}

\draw
  (2.6,+.7)
    node
  {$\xmapsto{
    \mbox{
      \small
      \rm
      topologize
    }
  }$};

\draw (0,-2.2) node
  {
    \footnotesize
    \begin{tabular}{c}
      \bf
      I. Classical computational trilogy
      \\
      \cite[\S 1]{Mellies06}\cite[\S 3]{Eades12}
    \end{tabular}
  };

\end{scope}

\begin{scope}[shift={(0,0)}]
\begin{scope}[rotate=(60), scale=(.7)]

\draw (0,0) circle (2.4);

\draw
  (45:2)
    node
  {
   \scalebox{1.2}{
   {
    \fcolorbox{black}{white}{\bf
    \hspace{-8pt}
    \color{orangeii}
    \begin{tabular}{c}
      Contextual Computation
      \\
      \& Programming Language
    \end{tabular}
    \hspace{-9pt}
    }}
  }};

\draw
  (45+120:2.4)
    node
  {
    \scalebox{1.2}{
    {
    \fcolorbox{black}{white}{\bf
    \hspace{-8pt}
    \color{darkblue}
    \begin{tabular}{c}
      Twisted Non-Abelian
      \\
      Cohomology Theory
    \end{tabular}
    \hspace{-9pt}
    }}
  }};

\draw
  (45+240:2.4)
    node
  {
   \scalebox{1.2}{
   {
    \fcolorbox{black}{white}{\bf
    \hspace{-8pt}
    \color{greenii}
    \begin{tabular}{c}
      Homotopy
      \\
      Type Theory
    \end{tabular}
    \hspace{-9pt}
    }
    }
  }};

\end{scope}

\draw (0,-2.2) node
  {
    \footnotesize
    \begin{tabular}{c}
      \bf
      II. Topological computational trilogy
      \\
      \cite{Shulman18}\cite[pp. 5-6]{SS20OrbifoldCohomology}
    \end{tabular}
  };

\end{scope}

\draw
  (2.5,+.7)
    node
  {$
    \xmapsto{
      \mbox{
        \small
        \rm
        quantize
      }
    }
  $};

\begin{scope}[shift={(5.3,0)}]
\begin{scope}[rotate=(60), scale=(.7)]

\draw (0,0) circle (2.4);

\draw
  (45:2)
    node
  {
    \scalebox{1.2}{
    {
    \fcolorbox{black}{white}{\bf
    \hspace{-8pt}
    \color{orangeii}
    \begin{tabular}{c}
      Classically Controlled
      \\
      Quantum Computation
    \end{tabular}
    \hspace{-9pt}
    }
    }
  }};

\draw
  (45+120:2.4)
    node
  {
   \scalebox{1.2}{
   {
    \fcolorbox{black}{white}{\bf
    \hspace{-8pt}
    \color{darkblue}
    \begin{tabular}{c}
      Twisted Generalized
      \\
      Cohomology Theory
    \end{tabular}
    \hspace{-9pt}
    }}
  }};

\draw
  (45+240:2.4)
    node
  {
    \scalebox{1.2}{
    {
    \fcolorbox{black}{white}{\bf
    \hspace{-8pt}
    \color{greenii}
    \begin{tabular}{c}
      Linear Homotopy
      \\
      Type Theory
    \end{tabular}
    \hspace{-9pt}
    }}
  }};

\end{scope}

\draw (0,-2.2) node
  {
    \footnotesize
    \begin{tabular}{c}
      \bf
     III. Quantum computational trilogy
      \\
      \cite{BaezStay11}\cite{Schreiber14}\cite[\S 4.1]{Schreiber13}
    \end{tabular}
  };
\end{scope}
\end{tikzpicture}
}

\vspace{-.05cm}

In particular, topological q-bits in
Hilbert spaces of anyon wavefunctions depending on
the classical position of the corresponding defects (vortices)
constitute the dependent linear data type which embody the
anyon's quantum states together with their adiabatic motion through
topological braid quantum gates (graphics taken form \cite{SS22AnyonictopologicalOrder}):

\vspace{-.45cm}
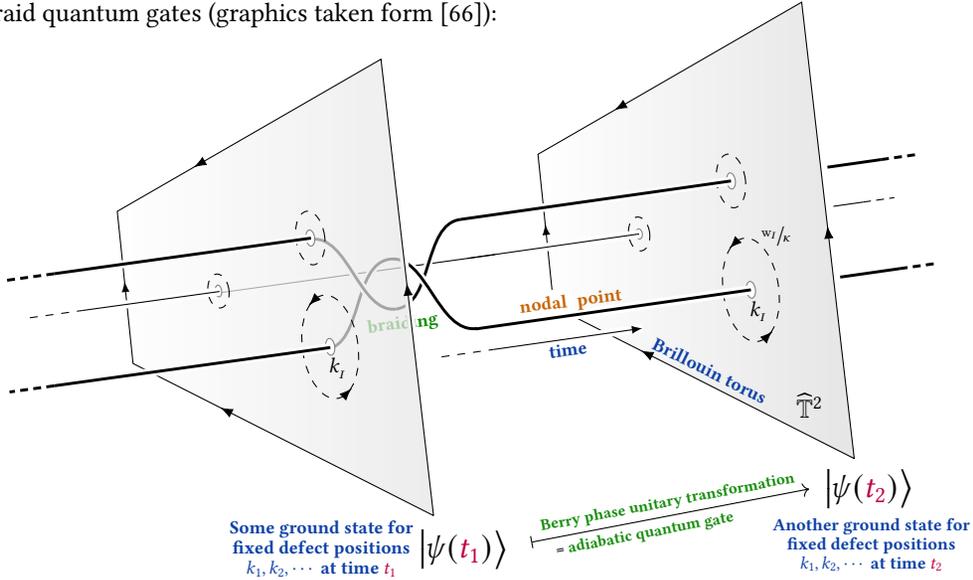
\begin{center}
\hypertarget{BraidGateSchematics}{}
\begin{tikzpicture}

  \shade[right color=lightgray, left color=white]
    (3,-3)
      --
      node[above, yshift=-1pt, sloped]{
        \scalebox{.7}{
          \color{darkblue}
          \bf
          Brillouin torus
        }
      }
    (-1,-1)
      --
    (-1.21,1)
      --
    (2.3,3);

  \draw[]
    (3,-3)
      --
    (-1,-1)
      --
    (-1.21,1)
      --
    (2.3,3)
      --
    (3,-3);

\draw[-Latex]
  ({-1 + (3+1)*.3},{-1+(-3+1)*.3})
    to
  ({-1 + (3+1)*.29},{-1+(-3+1)*.29});

\draw[-Latex]
    ({-1.21 + (2.3+1.21)*.3},{1+(3-1)*.3})
      --
    ({-1.21 + (2.3+1.21)*.29},{1+(3-1)*.29});

\draw[-Latex]
    ({2.3 + (3-2.3)*.5},{3+(-3-3)*.5})
      --
    ({2.3 + (3-2.3)*.49},{3+(-3-3)*.49});

\draw[-latex]
    ({-1 + (-1.21+1)*.53},{-1 + (1+1)*.53})
      --
    ({-1 + (-1.21+1)*.54},{-1 + (1+1)*.54});

  \begin{scope}[rotate=(+8)]
   \draw[dashed]
     (1.5,-1)
     ellipse
     ({.2*1.85} and {.37*1.85});
   \begin{scope}[
     shift={(1.5-.2,{-1+.37*1.85-.1})}
   ]
     \draw[->, -Latex]
       (0,0)
       to
       (180+37:0.01);
   \end{scope}
   \begin{scope}[
     shift={(1.5+.2,{-1-.37*1.85+.1})}
   ]
     \draw[->, -Latex]
       (0,0)
       to
       (+37:0.01);
   \end{scope}
   \begin{scope}[shift={(1.5,-1)}]
     \draw (.43,.65) node
     { \scalebox{.8}{$
       \sfrac{\weight_I}{\ShiftedLevel}
     $} };
  \end{scope}
  \draw[fill=white, draw=gray]
    (1.5,-1)
    ellipse
    ({.2*.3} and {.37*.3});
  \draw[line width=3.5, white]
   (1.5,-1)
   to
   (-2.2,-1);
  \draw[line width=1.1]
   (1.5,-1)
   to node[above, yshift=-3pt, pos=.85]{
     \;\;\;\;\;\;\;\;\;\;\;\;\;
     \rotatebox[origin=c]{7}
     {
     \scalebox{.7}{
     \color{orangeii}
     \bf
     \colorbox{white}{nodal} point
     }
     }
   }
   (-2.2,-1);
  \draw[
    line width=1.1
  ]
   (1.5+1.2,-1)
   to
   (3.5,-1);
  \draw[
    line width=1.1,
    densely dashed
  ]
   (3.5,-1)
   to
   (4,-1);

  \draw[line width=3, white]
   (-2,-1.3)
   to
   (0,-1.3);
  \draw[-latex]
   (-2,-1.3)
   to
   node[below, yshift=+3pt]{
     \scalebox{.7}{
       \rotatebox{+7}{
       \color{darkblue}
       \bf
       time
       }
     }
   }
   (0,-1.3);
  \draw[dashed]
   (-2.7,-1.3)
   to
   (-2,-1.3);

 \draw
   (-3.15,-.8)
   node{
     \scalebox{.7}{
       \rotatebox{+7}{
       \color{greenii}
       \bf
       braiding
       }
     }
   };

  \end{scope}

  \begin{scope}[shift={(-.2,1.4)}, scale=(.96)]
  \begin{scope}[rotate=(+8)]
  \draw[dashed]
    (1.5,-1)
    ellipse
    (.2 and .37);
  \draw[fill=white, draw=gray]
    (1.5,-1)
    ellipse
    ({.2*.3} and {.37*.3});
  \draw[line width=3.1, white]
   (1.5,-1)
   to
   (-2.3,-1);
  \draw[line width=1.1]
   (1.5,-1)
   to
   (-2.3,-1);
  \draw[line width=1.1]
   (1.5+1.35,-1)
   to
   (3.6,-1);
  \draw[
    line width=1.1,
    densely dashed
  ]
   (3.6,-1)
   to
   (4.1,-1);
  \end{scope}
  \end{scope}

  \begin{scope}[shift={(-1,.5)}, scale=(.7)]
  \begin{scope}[rotate=(+8)]
  \draw[dashed]
    (1.5,-1)
    ellipse
    (.2 and .32);
  \draw[fill=white, draw=gray]
    (1.5,-1)
    ellipse
    ({.2*.3} and {.32*.3});
  \draw[line width=3.1, white]
   (1.5,-1)
   to
   (-1.8,-1);
\draw
   (1.5,-1)
   to
   (-1.8,-1);
  \draw
    (5.23,-1)
    to
    (6.4-.6,-1);
  \draw[densely dashed]
    (6.4-.6,-1)
    to
    (6.4,-1);
  \end{scope}
  \end{scope}

\draw (2.41,-2.3) node
 {
  \scalebox{1}{
    $\DualTorus{2}$
  }
 };

\draw (1.73,-1.06) node
 {
  \scalebox{.8}{
    $k_{{}_{I}}$
  }
 };

\begin{scope}
[ shift={(-2,-.55)}, rotate=-82.2  ]

 \begin{scope}[shift={(0,-.15)}]

  \draw[]
    (-.2,.4)
    to
    (-.2,-2);

  \draw[
    white,
    line width=1.1+1.9
  ]
    (-.73,0)
    .. controls (-.73,-.5) and (+.73-.4,-.5) ..
    (+.73-.4,-1);
  \draw[
    line width=1.1
  ]
    (-.73+.01,0)
    .. controls (-.73+.01,-.5) and (+.73-.4,-.5) ..
    (+.73-.4,-1);

  \draw[
    white,
    line width=1.1+1.9
  ]
    (+.73-.1,0)
    .. controls (+.73,-.5) and (-.73+.4,-.5) ..
    (-.73+.4,-1);
  \draw[
    line width=1.1
  ]
    (+.73,0+.03)
    .. controls (+.73,-.5) and (-.73+.4,-.5) ..
    (-.73+.4,-1);

  \draw[
    line width=1.1+1.9,
    white
  ]
    (-.73+.4,-1)
    .. controls (-.73+.4,-1.5) and (+.73,-1.5) ..
    (+.73,-2);
  \draw[
    line width=1.1
  ]
    (-.73+.4,-1)
    .. controls (-.73+.4,-1.5) and (+.73,-1.5) ..
    (+.73,-2);

  \draw[
    white,
    line width=1.1+1.9
  ]
    (+.73-.4,-1)
    .. controls (+.73-.4,-1.5) and (-.73,-1.5) ..
    (-.73,-2);
  \draw[
    line width=1.1
  ]
    (+.73-.4,-1)
    .. controls (+.73-.4,-1.5) and (-.73,-1.5) ..
    (-.73,-2);

 \draw
   (-.2,-3.3)
   to
   (-.2,-2);
 \draw[
   line width=1.1,
   densely dashed
 ]
   (-.73,-2)
   to
   (-.73,-2.5);
 \draw[
   line width=1.1,
   densely dashed
 ]
   (+.73,-2)
   to
   (+.73,-2.5);

  \end{scope}
\end{scope}

\begin{scope}[shift={(-5.6,-.75)}]

  \draw[line width=3pt, white]
    (3,-3)
      --
    (-1,-1)
      --
    (-1.21,1)
      --
    (2.3,3)
      --
    (3, -3);

  \shade[right color=lightgray, left color=white, fill opacity=.7]
    (3,-3)
      --
    (-1,-1)
      --
    (-1.21,1)
      --
    (2.3,3);

  \draw[]
    (3,-3)
      --
    (-1,-1)
      --
    (-1.21,1)
      --
    (2.3,3)
      --
    (3, -3);

\draw (1.73,-1.06) node
 {
  \scalebox{.8}{
    $k_{{}_{I}}$
  }
 };

\draw[-Latex]
  ({-1 + (3+1)*.3},{-1+(-3+1)*.3})
    to
  ({-1 + (3+1)*.29},{-1+(-3+1)*.29});

\draw[-Latex]
    ({-1.21 + (2.3+1.21)*.3},{1+(3-1)*.3})
      --
    ({-1.21 + (2.3+1.21)*.29},{1+(3-1)*.29});

\draw[-Latex]
    ({2.3 + (3-2.3)*.5},{3+(-3-3)*.5})
      --
    ({2.3 + (3-2.3)*.49},{3+(-3-3)*.49});

\draw[-latex]
    ({-1 + (-1.21+1)*.53},{-1 + (1+1)*.53})
      --
    ({-1 + (-1.21+1)*.54},{-1 + (1+1)*.54});

  \begin{scope}[rotate=(+8)]
   \draw[dashed]
     (1.5,-1)
     ellipse
     ({.2*1.85} and {.37*1.85});
   \begin{scope}[
     shift={(1.5-.2,{-1+.37*1.85-.1})}
   ]
     \draw[->, -Latex]
       (0,0)
       to
       (180+37:0.01);
   \end{scope}
   \begin{scope}[
     shift={(1.5+.2,{-1-.37*1.85+.1})}
   ]
     \draw[->, -Latex]
       (0,0)
       to
       (+37:0.01);
   \end{scope}
  \draw[fill=white, draw=gray]
    (1.5,-1)
    ellipse
    ({.2*.3} and {.37*.3});
 \end{scope}

   \begin{scope}[shift={(-.2,1.4)}, scale=(.96)]
  \begin{scope}[rotate=(+8)]
  \draw[dashed]
    (1.5,-1)
    ellipse
    (.2 and .37);
  \draw[fill=white, draw=gray]
    (1.5,-1)
    ellipse
    ({.2*.3} and {.37*.3});
\end{scope}
\end{scope}

  \begin{scope}[shift={(-1,.5)}, scale=(.7)]
  \begin{scope}[rotate=(+8)]
  \draw[dashed]
    (1.5,-1)
    ellipse
    (.2 and .32);
  \draw[fill=white, draw=gray]
    (1.5,-1)
    ellipse
    ({.2*.3} and {.37*.3});
\end{scope}
\end{scope}

\begin{scope}
[ shift={(-2,-.55)}, rotate=-82.2  ]

 \begin{scope}[shift={(0,-.15)}]

 \draw[line width=3, white]
   (-.2,-.2)
   to
   (-.2,2.35);
 \draw
   (-.2,.5)
   to
   (-.2,2.35);
 \draw[dashed]
   (-.2,-.2)
   to
   (-.2,.5);

\end{scope}
\end{scope}

\begin{scope}
[ shift={(-2,-.55)}, rotate=-82.2  ]

 \begin{scope}[shift={(0,-.15)}]

 \draw[
   line width=3, white
 ]
   (-.73,-.5)
   to
   (-.73,3.65);
 \draw[
   line width=1.1
 ]
   (-.73,.2)
   to
   (-.73,3.65);
 \draw[
   line width=1.1,
   densely dashed
 ]
   (-.73,.2)
   to
   (-.73,-.5);
 \end{scope}
 \end{scope}

\begin{scope}
[ shift={(-2,-.55)}, rotate=-82.2  ]

 \begin{scope}[shift={(0,-.15)}]

 \draw[
   line width=3.2,
   white]
   (+.73,-.6)
   to
   (+.73,+3.7);
 \draw[
   line width=1.1,
   densely dashed]
   (+.73,-0)
   to
   (+.73,+-.6);
 \draw[
   line width=1.1 ]
   (+.73,-0)
   to
   (+.73,+3.71);
\end{scope}
\end{scope}

\end{scope}

\draw
  (-2.2,-4.2) node
  {
    \scalebox{1.2}{
      $
       \mathllap{
          \raisebox{1pt}{
            \scalebox{.58}{
              \color{darkblue}
              \bf
              \def\arraystretch{.9}
              \begin{tabular}{c}
                Some ground state for
                \\
                fixed defect positions
                \\
                $k_1, k_2, \cdots$
                at time
                {\color{purple}$t_1$}
              \end{tabular}
            }
          }
          \hspace{-5pt}
       }
        \big\vert
          \psi({\color{purple}t_1})
        \big\rangle
      $
    }
  };

\draw[|->]
  (-1.3,-4.1)
  to
  node[
    sloped,
    yshift=5pt
  ]{
    \scalebox{.6}{
      \color{greenii}
      \bf
      Berry phase
      unitary transformation
    }
  }
  node[
    sloped,
    yshift=-5pt,
    pos=.4
  ]{
    \scalebox{.6}{
      \color{greenii}
      \bf
      {\color{black}=}
      adiabatic quantum gate
      }
  }
  (+2.4,-3.4);

\draw
  (+3.2,-3.85) node
  {
    \scalebox{1.2}{
      $
        \underset{
          \raisebox{-7pt}{
            \scalebox{.55}{
              \color{darkblue}
              \bf
              \def\arraystretch{.9}
               \begin{tabular}{c}
              Another ground state for
                \\
                fixed defect positions
                \\
                $k_1, k_2, \cdots$
                at time
                {\color{purple}$t_2$}
              \end{tabular}
            }
          }
        }{
        \big\vert
          \psi({\color{purple}t_2})
        \big\rangle
        }
      $
    }
  };

\end{tikzpicture}
\end{center}
\vspace{-.25cm}

Notice how the full set of language features of {\it dependent linear data types} in {\it cohesive homotopy type theory}  is necessary and sufficient for faithfully capturing this state of affairs.
We  now close this note by indicating\footnote{For details see the supplementary material avialable at: \href{https://ncatlab.org/schreiber/show/Topological+Quantum+Computation+in+TED-K\#GMConAbs}{\tt ncatlab.org/schreiber/show/TQCinTEDK\#GMConAbs}} the syntax of {\tt TED-K} in cohesive HoTT which provides the programming language reflection of such topological quantum gates. This is a research project at the {\it Center for Quantum and Topological Systems} (\href{https://ncatlab.org/nlab/show/Center+for+Quantum+and+Topological+Systems}{\underline{CQTS}}), whose detailed results will be reported elsewhere.

\newpage

\noindent
{\bf The language {\tt TED-K}.}
In cohesive homotopy type theory one has access to the data type $\mathbb{R}^1$ of real numbers understood {\it with} its differential topology \cite{Shulman15}.
From this, one constructs data types representing familiar objects in differential topology and analysis, such as separable complex Hilbert spaces $\mathscr{H}$, their projective unitary groups $\mathrm{PU}$ and their spaces of Fredholm operators $\FredholmOperators^0_{\ComplexNumbers}$, etc., all understood with their respective (strong operator) topology (see \cite[Ex. 1.3.19]{SS21EPB} for details and further pointers).

This way, complex topological K-theory groups of any cohesive data type $\mathcal{X}$ are coded as the type of cohesive homotopy classes of Fredholm-operator valued functions:
\begin{equation}
  \label{KTheoryCodedInTypeTheory}
  \mathcal{X}
  \,\colon\,
  \mathrm{Type}
  \quad
  \vdash
  \quad
  \mathrm{KU}
  \big(\mathcal{X}\big)
  \;\equiv\;
  \big\vert
  \,
  \shape\big(
    \mathcal{X}
    \xrightarrow{\;}
    \FredholmOperators^0_{\ComplexNumbers}
  \big)
  \big\vert_0
  \,\colon\,
  \mathrm{Type}
  \,.
\end{equation}
Here ``\shape'' denotes the {\it shape modality} operator of cohesive HoTT and $\vert-\vert_0$ denotes the 0-truncation modality present already in plain HoTT (see \cite{RSS20}).

Moreover, the type of (self-adjoint odd-graded) Fredholm operators
receives a canonical conjugation action by the (graded) projective unitary group $\mathrm{PU}$ and hence (by \cite{NSS12}\cite{BvDR18}, see \cite[Prop. 0.2.1]{SS21EPB})
occurs as the homotopy fiber of a dependent type $\HomotopyQuotient{\FredholmOperators^0_{\ComplexNumbers}}{\mathrm{PU}}$ over the {\it delooping type} $\mathbf{B}\mathrm{PU}$. Regarding a function from any type $\mathcal{X}$ to $\mathbf{B}\mathrm{PU}$ as a twist, we obtain the corresponding {\it twisted K-cohomology} type as the 0-truncation of the shape of the dependent product of the function type:
\begin{equation}
  \label{TwistedKTheoryCodedInTypeTheory}
  \mathcal{X}
  \,\colon\,
  \mathrm{Type}
  ;\,
  \tau
  \,\colon\,
  \mathcal{X}
  \xrightarrow{\;}
  \mathbf{B}\mathrm{PU}
  \quad
  \vdash
  \quad
  \mathrm{KU}^\tau
  \big(\mathcal{X}\big)
  \,\equiv\,
  \bigg\vert
  \,
    \shape \,
    {\prod}_{\mathbf{B}\mathrm{PU}}
    \Big(
      \mathcal{X}
      \xrightarrow{\;}
      \HomotopyQuotient{
        \FredholmOperators^0_{\ComplexNumbers}
        }
        { \mathrm{PU} }
    \Big)
  \bigg\vert_0
  \;:\;
  \mathrm{Type}
\end{equation}
If $\mathcal{X}$ itself here is not 0-truncated, then this is {\it twisted orbifold K-theory}. Specifically, if $\mathcal{X} \,\simeq\, \HomotopyQuotient{\TopologicalSpace}{G}$ for a discrete group $G$ acting on a 0-truncated type $\mathrm{X}$, then this is {\it twisted $G$-equivariant K-theory}.

By this we mean that under interpreting homotopy type theory into $\infty$-topoi (reviewed in \cite{Riehl22}), and here specifically into the cohesive $\infty$-topos of smooth $\infty$-groupoids (\cite{Schreiber13}, consise details in \cite[Prop. A.56]{FSS20Character}\cite[\S 3.3.1]{SS21EPB}), the above types are identified with the twisted equivariant K-theory groups found in the traditional literature (\cite{AtiyahSegal04}\cite{LupercioUribe01}\cite{AdemRuan01}, see \cite[Ex. 4.3.19]{SS21EPB}).

\medskip

\noindent
{\bf Topological quantum syntax.}
The key point now is  this \cite{SS22AnyonictopologicalOrder}: Taking the type $\mathrm{X}$  to be a configuration space of points in the torus, then these twisted equivariant K-groups
\eqref{TwistedKTheoryCodedInTypeTheory}
naturally subsume the (mass terms among) spaces of topologically ordered $\mathfrak{su}(2)$-anyonic ground states (\cite[Thm. 2.18]{SS22AnyonicDefectBranes}), and they do so as linear types {\it dependent} on the classical/external/intuitionistic parameter of anyonic defect positions:
\begin{equation}
  \label{AnyonsCodedInTypeTheory}
  \underset{
    \mathrlap{
    \hspace{-1.8cm}
    \raisebox{-.5pt}{
      \tiny
      \color{orangeii}
      \bf
      Defect anyon positions
    }
    }
  }{
  \overset{
    \mathrlap{
    \hspace{-1.8cm}
    \raisebox{2pt}{
      \tiny
      \color{darkblue}
      \bf
      Term of base type
    }
    }
  }{
  \{
    k_I
  \}_{I=1}^N
  \,:\,
  \HomotopyQuotient
  {
  \ConfigurationSpace{N}
  \big(
    \DualTorus{2}
  \big)
  }{
    G
  }
  }
  }
  \quad
  \vdash
  \quad
  \underset{
    \mathrlap{
    \hspace{-5.1cm}
    \raisebox{-1pt}{
       {  \tiny
      \color{greenii}
      \bf  parameterize}
      \;
     {
      \tiny
      \color{orangeii}
      \bf Hilbert space of
      topologically ordered ground state wavefunctions
      }
    }
    }
  }{
  \overset{
    \mathrlap{
    \hspace{-4.65cm}
    \raisebox{0pt}{
     {\color{greenii}     \tiny \bf entails}
      \; \quad
 {\tiny
      \color{darkblue}
      \bf      dependent linear TED-K type}
    }
    }
  }{
  \Big\vert
  \,
  \shape
  \,
  {\prod}_{\mathbf{B}\mathrm{PU}}
  \Big(
    \HomotopyQuotient{
    \ConfigurationSpace{n}
    \big(
      \DualTorus{2}
        \setminus
      \{k_I\}_{I=1}^M
    \big)
    }{G}
    \xrightarrow{\;}
    \HomotopyQuotient
      { \FredholmOperators^0_{\ComplexNumbers} }
      { \mathrm{PU} }
  \Big)
  \Big\vert_0
  }
  }
  :
  \mathrm{Type}
\end{equation}
This turns out to depend only on the cohesive {\it shape} $\shape$ of the configuration space -- which is equivalently (the delooping of) the (toroidal) braid group (e.g. \cite[\S 2.1]{Kohno02}\cite[\S 8]{EtingofFrenkelKirillov98}):
\begin{equation}
  \label{ShapeOfConfigurationSpaceOfPoints}
  \mathllap{
    \mbox{
      \tiny
      \color{darkblue}
      \bf
      \def\arraystretch{.9}
      \begin{tabular}{c}
        Cohesive shape
        \\
        of base type
      \end{tabular}
    }
  }
  \shape
  \,
  \Big(
    \ConfigurationSpace{N}
    \big(
      \DualTorus{2}
    \big)
  \Big)
  \;\simeq\;
  \mathbf{B}
  \big(
  \mathrm{Br}_{{}_{\DualTorus{2}}}(N)
  \big)
  \mathrlap{
    \tiny
    \color{orangeii}
    \bf
    \def\arraystretch{.9}
    \begin{tabular}{c}
  \bf    Delooping stack
      \\
   \bf   of braid group
      {\color{black}.}
    \end{tabular}
  }
\end{equation}
Therefore, the ambient univalent homotopy type theory now provides the operation of {\it transport} (\cite[\S 2.3]{UFP13}) of the dependent type \eqref{AnyonsCodedInTypeTheory} along identities in the cohesive shape of its base type \eqref{ShapeOfConfigurationSpaceOfPoints}:
$$
  \underset{
    \mathclap{
      \hspace{-.5cm}
      \raisebox{-0pt}{
        \tiny
        \color{orangeii}
        \bf
        Braid of defect anyon worldlines
      }
    }
  }{
  \overset{
    \mathclap{
    \hspace{-12pt}
    \raisebox{2pt}{
      \tiny
      \color{darkblue}
      \bf
      \begin{tabular}{c}
        Term in identity type of base type
      \end{tabular}
    }
    }
  }{
  \gamma \,:
  \bigg(
  (k_I)_{I =1}^N
  \underset
  {
    \mathclap{
    \raisebox{-6pt}{
    \scalebox{.7}{$
      \HomotopyQuotient{
      \shape \mathrm{Conf}_N\big(\DualTorus{2}\big)
      }
      {G}
    $}
    }
    }
  }{
    \;\;
    =
    \;\;
  }
  (k_I)_{I =1}^N
  \bigg)
  }
  }
  \;\;
  \vdash
  \;\;
  \underset{
    \mathrlap{
    \hspace{-3.35cm}
    \raisebox{-6.2pt}{
          {\color{greenii}     \tiny \bf  induces} \;
    {\tiny
      \color{orangeii}
      \bf  unitary quantum gate operation
      on topological q-bits}
      }
    }
  }{
  \overset{
    \mathrlap{
    \hspace{-3.35cm}
    \raisebox{5.5pt}{
          {\color{greenii}     \tiny \bf   induces}
      \;
      {\tiny
      \color{darkblue}
      \bf dependent type transport operation on TED-K types
      }
    }
    }
  }{
  U(\gamma)
  \,:\,
  \mathrm{KU}^\tau_G
  \Big(
    \ConfigurationSpace{n}
    \big(
      \DualTorus{2}
      \setminus
      (k_I)_{I=1}^N
    \big)
  \Big)
  }
  }
  \xrightarrow{\;}
  \mathrm{KU}^\tau_G
  \Big(
    \ConfigurationSpace{n}
    \big(
      \DualTorus{2}
      \setminus
      (k_I)_{I=1}^N
    \big)
  \Big).
$$
By the above discussion, this term denotes an operation of the braid group on the space of anyon ground states and, as such, it encodes the desired braid quantum gates as indicated on \hyperlink{BraidGateSchematics}{p. 4}.

\end{document}